# Between Regulation and Accessibility: How Chinese University Students Navigate Global and Domestic Generative AI

Qin Xie, University of Minnesota, USA

Ming Li, The University of Osaka, Japan

Fei Cheng, Kyoto University, Japan

## Abstract

Despite the rapid proliferation of generative AI in higher education, students in China face significant barriers in accessing global tools like ChatGPT due to regulations and constraints. Grounded in the Unified Theory of Acceptance and Use of Technology 2 (UTAUT2) model, this study employs qualitative interviews to investigate how Chinese university students interact with both global and domestic generative AIs in the learning process. Findings reveal that engagement is shaped by accessibility, language proficiency, and cultural relevance. Students often employ workarounds (e.g., VPNs) to access global generative AIs, raising ethical and privacy concerns. Domestic generative AIs, while offering language and cultural advantages, are limited by content filtering and output constraints. This research contributes to understanding generative AI adoption in non-Western contexts by highlighting the complex interplay of political, linguistic, and cultural factors. It advocates for human-centered, multilingual, domestic context-sensitive AI integration to ensure equitable and inclusive digital learning environments.

**Introduction**

With the advent of ChatGPT, generative AI has been rapidly adopted in higher education, profoundly transforming students' learning approaches and academic outcomes (Rasul et al., 2023). The integration of generative AI into higher education provides students with more efficient learning assistants, supporting autonomous learning, enhancing writing and research capabilities, and facilitating personalized education (Baidoo-Anu & Ansah, 2023; Li et al., 2025a; Udeh, 2025). However, it also raises concerns regarding academic integrity, the reliability of generated information, and students' over-reliance on AI-generated content (Li et al., 2024; Sullivan & McLaughlan, 2023; Yusuf & Román-González, 2024).

In response to these opportunities and concerns, efforts at the global, national, and institutional levels have emerged to regulate the use of generative AI in academic settings. Organizations such as the United Nations Educational, Scientific and Cultural Organization (UNESCO) and the European Union (EU) have issued guidelines to help universities regulate the use of generative AI in academic settings (European Commission, 2023; Miao & Holmes, 2023). While a few countries have implemented policies to address the implications of generative AI in education, others remain in a phase of observation or exploration (Vidal et al., 2023; Xie et al., 2024). Leading universities worldwide have begun developing institutional policies and faculty training programs aimed at fostering the ethical use of generative AI in higher education (Dabis & Csáki, 2024). However, national governments and universities exhibit varying levels of responsiveness and regulatory emphasis, resulting in significant disparities in students' access to policy frameworks and technical support when adopting generative AI (Li et al., 2025b; Wang et al., 2024).

Moreover, the development of mainstream generative AI models remains predominantly English-centric and often reflects Western cultural norms (Bender et al., 2021; Liu et al., 2025; Zhong et a., 2024). This linguistic and cultural orientation poses challenges for students in non-English-speaking countries, particularly for users of low-resource languages who may experience reduced model performance due to limited training data (Sun et al., 2024). For example, languages spoken in regions such as Southeast Asia and Africa continue to demonstrate weaker performance in existing models, reinforcing digital linguistic inequalities (Aji et al., 2023; Kshetri, 2024). Additionally, global economic disparities contribute to an AI digital divide, wherein students in developed countries have greater access to premium generative AI (e.g., ChatGPT Plus, Claude Pro), while those in developing or less affluent regions face challenges related to accessibility, infrastructure, and educational support.

Access to generative AI is further complicated by regulatory differences across countries. In nations such as China, Russia, Iran, and North Korea, varying degrees of restrictions on tools like ChatGPT limit students' access to internationally developed generative AI platforms (Open AI, n.d.). At the same time, the global generative AI landscape is largely dominated by the United States (Brookings Institution, 2023). In response, many countries are actively pursuing the development of domestic AI alternatives. China, in particular, has emerged as a significant case: due to policy restrictions on accessing foreign AI platforms such as ChatGPT, Claude, and Gemini, the country has accelerated the growth of its own generative AI technologies (Smith, 2025, January 28). A number of homegrown alternatives have gained traction, including Kimi Chat (月之暗面), ChatGLM (智谱清言), Doubao (豆包), Wenxin Yiyan (文心一言), and Qwen (通义千问). These domestic generative AIs are continuously evolving, with improvements in Chinese language processing capabilities, industry applications, and user experience, making them viable generative AI options for Chinese students in their daily learning and research.

Against this unique and evolving digital landscape, this study seeks to explore: How do Chinese undergraduate students engage with and navigate global and domestic generative AI in their learning process? Specifically, it aims to examine their patterns of selection, strategies for adaptation, perceived benefits, and encountered challenges across different academic scenarios.

## Literature review

### The Technical Overview of Global Generative AI and Domestic Generative AI

Generative AI has attracted worldwide attention since OpenAI launched ChatGPT in late 2022. ChatGPT, a large language model powered by deep learning, can generate high-quality responses to assist humans in a wide range of downstream tasks (Brown et al., 2020). In fact, OpenAI had anticipated the Scaling Law for generative AI models since 2020 (Kaplan et al. 2020), predicting that model performance would grow predictably with increases in parameters, computational resources, and training data. The emergence of the Scaling Law immediately spurred a global race among major tech companies —including Google, Microsoft, Meta, and NVIDIA— to scale generative AI (Chowdhery et al., 2023; Shoeybi et al., 2019; Smith et al., 2022; Touvron et al., 2023).

China's generative AI development finds itself in a paradoxical position due to the geopolitical tensions between the U.S. and China. The Chinese government has banned global generative AI services such as Gemini, Claude, and Copilot (Cyberspace Administration of China, 2023). This creates a gap in user demand that is inevitably filled by domestic generative AI products. Meanwhile, Chinese tech companies face hurdles in developing domestic generative AI models. The U.S. government has led to export restrictions on high-end NVIDIA chips to China (Feng, 2025), while domestic alternatives struggle to compete with CUDA (Compute Unified Device Architecture), a well-established software ecosystem of Nvidia. Consequently, Chinese companies have mainly focused on small- to mid-scale models, such as Alibaba's Qwen series, which ranges from 7B to 70B parameters. However, the landscape began to shift in late 2024. Efficiency-driven techniques like Mixture of Experts (Jiang et al., 2023) significantly reduce reliance on computational resources. A notable example is DeepSeek, which optimized efficiency-driven techniques to achieve ChatGPT-level performance—despite limitations in computing resources (Guo et al., 2025; Liu et al., 2024).

**Technical Comparison of Global and Domestic Generative AI**

***Generative AI Platforms and Multiple Modalities***

In this study, global AI platforms refer primarily to artificial intelligence systems developed and operated by leading international technology companies. These platforms are characterized by globally-oriented technological architectures, broad user bases, and international influence. A representative example is OpenAI's ChatGPT, which is trained on diverse datasets, designed for global markets, and accessible in multiple countries and regions. Such platforms typically use English as the primary language and exhibit strong general capabilities, representing the forefront of global AI development.

In contrast, domestic AI platforms refer to AI systems developed and deployed by Chinese companies, primarily serving the domestic market. Examples include Deepseek, Doubao, and Qwen. These platforms are specifically adapted for the Chinese context through algorithm optimization, domesticized training data, advanced Chinese language processing, and compliance with domestic regulatory and cultural requirements.

Generative AI platforms are generally offered as online services to subscription users, with their backbone models continuously evolving. For example, when OpenAI first launched ChatGPT, its base model was GPT-3 (Brown et al., 2020). It has been upgraded to GPT-4o since May 2024, with even more cutting-edge versions currently in testing (Achiam et al., 2023). We separately sampled three global (GPT-4o, Claude-3.5, Gemini-Exp) and three Chinese (Deepseek-v3, Doubao-1.5-Pro, Qwen2.5) generative AI products for comparison based on their performance, release date, and openness of technical details.

Beyond advancements in language generation (text-to-text), generative AI has increasingly expanded its integration with the visual modality (Esser et al., 2024; Rombach et al., 2022;). Text-to-image models, such as Midjourney, leverage generative AIs' deep understanding of users' textual instructions to generate desired images. This study covers the usage of both text-to-text and text-to-image models. However, due to the unstandardized evaluation for image generation, the technical comparisons in later this study will primarily focus on text-to-text models.

### *Assessment Dimension*

Generative AI possesses generalized problem-solving capabilities, making its comprehensive evaluation particularly challenging, as existing benchmarks are often limited to specific dimensions. Over the past two years, the AI community has invested significant effort in developing rigorous evaluations to assess model performance across multiple dimensions better (Chang et al., 2024). For instance, MMLU (Massive Multitask Language Understanding) is widely used to evaluate a model's knowledge base and problem-solving ability in English. It covers 57 subjects across STEM, humanities, and social sciences, with difficulty levels ranging from elementary to professional (Hendrycks et al., 2020). Inspired by MMLU's multi-subject and multi-difficulty design, the Chinese dataset C-Eval was developed as an equivalent benchmark for evaluating models' Chinese ability (Huang et al., 2023). Beyond linguistic proficiency, generative AI has demonstrated the ability to generate human-quality responses but struggles with math calculations. Recent efforts have focused on math, reasoning, and coding tasks for measuring generative AIs' complex problem-solving capabilities (Austin et al., 2021; Hendrycks et al., 2021; Suzgun et al., 2022). In this study, we define five key dimensions for comparing global and domestic generative AI models: **English**, **Chinese**, **math**, **reasoning**, and **coding**. For each dimension, we select a representative benchmark to conduct a rigorous comparative analysis.

Figure 1 presents a radar chart illustrating the quantitative comparison across five evaluation dimensions. The solid lines represent global models, while the dash-dot lines denote domestic generative AI models. Key observations from the results include:

- **Language capabilities:** English proficiency is similar across models, likely due to the abundance of English resources for training. However, domestic models consistently outperform global models in Chinese C-Eval, benefiting from their dedicated high-quality Chinese corpora.

- **Reasoning and Coding capabilities:** In the reasoning and coding evaluation, the performance of global and domestic generative AI models are relatively close.

- **Math capability:** Recent advancements in generative AI have heavily focused on improving math reasoning. Consequently, models released more recently tend to exhibit significantly stronger math capability, such as Gemini-Exp, DeepSeek-v3, and Doubao-1.5-Pro.

**Figure 1**

*Performance Comparison Between Global (solid lines) and Domestic Generative AI (dash-dot lines) Across Five Dimensions*

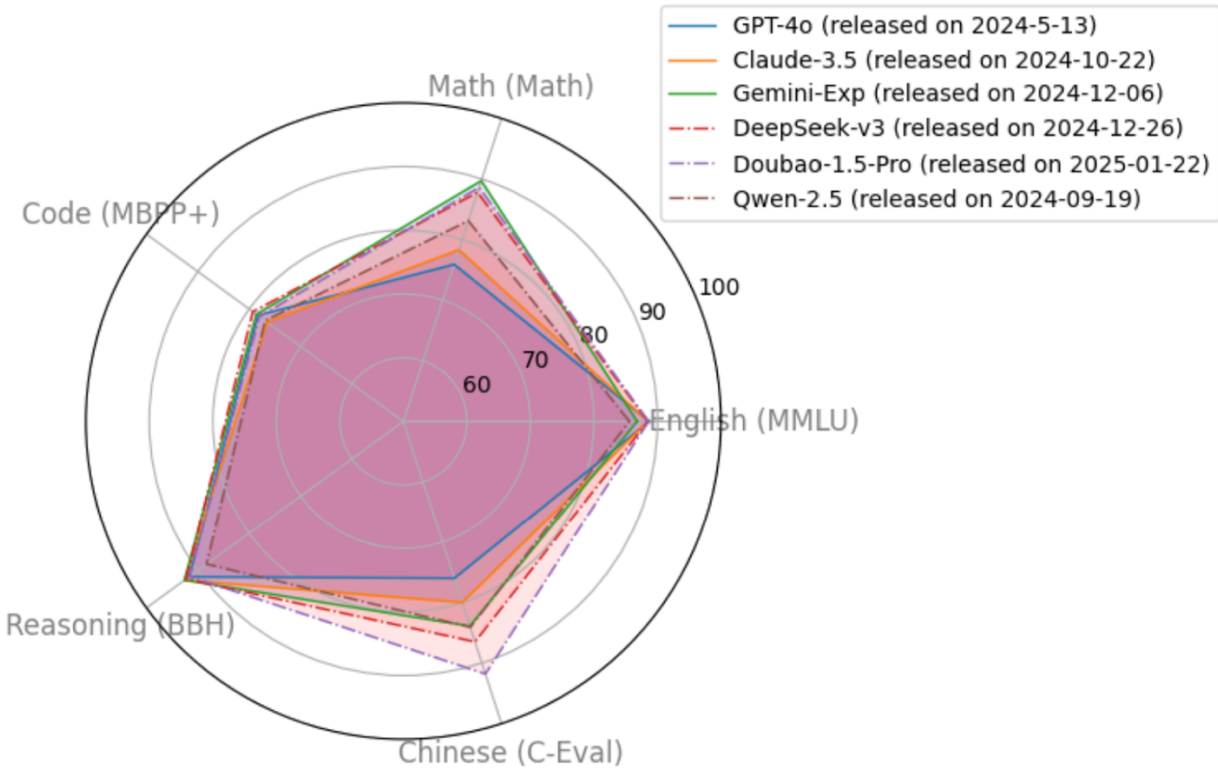

*Note.* The scores are collected from publicly available sources, including research papers and official product pages, with efforts made to verify consistency across different sources.

In summary, there is no significant performance gap between global and domestic models at present. Companies in both the U.S. and China are actively conducting cutting-edge research, continuously enhancing the capabilities of generative AI. This is also reflected in academic achievements. The U.S. and China consistently lead in the number of accepted papers at top-tier international generative AI conferences, significantly outpacing other countries in research contributions (ACL Rolling Review Statistics, 2025).

## Theoretical Framework

The Unified Theory of Acceptance and Use of Technology (UTAUT) and its extension model UTAUT2 (Venkatesh et al., 2003; Venkatesh et al., 2012) are the widely used theoretical frameworks to explain user intentions to adopt and use technology (Xia & Chen, 2024). Previously, they have some explanatory models of individual acceptance of information technology in psychology and sociology (Ayaz & Yanartaş, 2020). UTAUT synthesizes several earlier models, including the Theory of Reasoned Action (Davis, 1989), Technology Acceptance Model (Mathieson, 1991), Motivation Model (Vallerand & Zanna, 1997), Theory of Planned Behavior (Ajzen, 1991), Unified Model of Technology Acceptance and Planned Behavior (Taylor & Todd, 1995), PC Utilization Model (Triandis, 1977), Diffusion Theory (Rogers, 1995), and Social Cognitive Theory (Bandura, 1986). Through an analysis of the strengths and limitations of eight theoretical frameworks, UTAUT integrates three key determinants that directly influence the intention to use technology: *Performance Expectancy*, *Effort Expectancy*, *Social Influence* and *Facilitating Conditions*. Furthermore, *Experience*, *Voluntariness of Use*, *Gender,* and *Age* are identified as significant moderating variables within the UTAUT model (Venkatesh et al., 2003).

Since its initial release, UTAUT has served as a foundational model and has been applied to research on various technologies in both organizational and non-organizational contexts (Ursavaş, 2022). With advancements in technology, Venkatesh et al. (2012) expanded the model by introducing three new constructs, *Hedonic Motivation*, *Price Value*, and *Habit* resulting in the development of UTAUT2. Therefore, the UTAUT2 model is built on seven key factors and three key moderators to predict users' technology use behavior and behavioral intention.

To be more detailed, according to Venkatesh et al. (2012), *Performance Expectancy* refers to the extent to which using a technology is expected to offer benefits to users in carrying out specific tasks. *Effort Expectancy* refers to the level of ease experienced by consumers when using technology. *Social Influence* refers to the degree to which individuals feel that important people in their lives, such as family and friends, think they should use a particular technology. *Facilitating Conditions*, relates to users' perceptions of the resources and support available to help them effectively use the technology (Venkatesh et al., 2003). *Hedonic Motivation* refers to the enjoyment or satisfaction individuals experience when using a technology (Brown & Venkatesh, 2005). *Price Value* is defined as the mental evaluation consumers make when weighing the perceived benefits of an application against the financial cost of using it (Dodds et al., 1991). *Habit* refers to a perceptual construct that reflects the influence of past experiences (Venkatesh et al., 2012).

In addition, *Age, Gender* and *Experience* are hypothesized to moderate the effects of seven constructs on behavioral intention and technology use (Venkatesh et al., 2003; Venkatesh et al., 2012). In this study, we will focus more on gender and experience, as all participants are of a similar age, in their early 20s.

UTAUT and its extended model, UTAUT2, have been applied across three broad categories within the context of generative AI in education. The first category involves applying these models to the teaching and learning process (Zhang & Wareewanich, 2024; Patterson et al., 2024). The second focuses on their application across different educational levels (Zaim et al., 2024; Du & Lv, 2024). The third category examines the adaptation of these models in diverse geographical and cultural contexts (Cao & Peng, 2024; Abdalla, 2025).

## Methodology

### Semi-structured Interview

Qualitative interviewing, often described as the art of listening to data, is a powerful method for exploring complex and evolving social processes (Rubin & Rubin, 2011). Grounded in conversational exchange, it highlights the dynamic interaction between interviewer and participant, with the researcher actively posing questions and listening closely to responses (Kvale, 1996; Warren, 2002). Among the various approaches, semi-structured interviews are most commonly used for their balance of structure and flexibility (Kallio et al., 2016). This study employs semi-structured interviews to examine Chinese undergraduate students' experiences and perspectives on integrating generative AI into their learning. The interview protocol includes both open- and closed-ended questions across five sections: academic background, generative AI usage, institutional guidance and support, attitudes toward generative AI, and reflections. The questions aim to understand how students currently use both global and domestic generative AI.

**Sampling**

The study population consists of Chinese-speaking undergraduate students enrolled in higher education institutions in China. Participants were recruited through WeChat, a widely used Chinese social media platform, where a recruitment announcement was posted. By leveraging the researchers' personal network, 15 participants were successfully recruited, including 8 female students and 7 male students, with 6 from STEM fields and 9 from the social sciences and humanities. All participants are students at one of China's most prestigious universities, located in a first-tier city. As a token of appreciation, each participant received a voucher for their involvement.

**Data Collection**

The interviews were conducted by the authors via the Zoom platform between July and August 2024. Each interview lasted approximately 60 minutes. At the start of each session, the researcher read the consent form, which outlined the study's objectives, details on data usage, and an overview of the interview questions. Participants were given the opportunity to ask questions and freely decide whether to participate. With the participants' consent, the interviews were recorded, and the discussions were automatically transcribed using Zoom's transcription feature.

These automated transcripts were then carefully reviewed, manually corrected for accuracy, and supplemented with relevant contextual notes. To ensure confidentiality, all identifying information was removed and pseudonyms were systematically assigned to each participant throughout the study.

**Data Analysis**

All transcripts were imported into the qualitative data analysis (QDA) software NVivo 14 Mac in chronological order. The authors conducted the coding process, applying Saldaña's (2025) two-cycle coding methods. In the first cycle, the data were broken down into manageable segments using in vivo coding and descriptive coding to capture key concepts and summarize content. Following this, the second cycle focused on identifying broader themes and conceptual patterns through pattern coding, focused coding, and axial coding (Saldaña, 2025). Overall, the transcripts were analyzed thematically, employing a combination of deductive and inductive coding approaches to uncover both predefined and emerging themes.

<div align="center">

**Results**

</div>

**Accessibility as a Facilitating Condition: Constraints and Workarounds**

*Constraints in Accessibility*

The accessibility of generative AI tools has become a key factor influencing students' usage patterns. Under the framework of Facilitating Conditions (Venkatesh et al., 2012), it is evident that government restrictions on global AI platforms such as ChatGPT, along with the availability of domestic alternatives, have significantly shaped students' behavior. ChatGPT and some other global generative AI are inaccessible in China due to the government's stringent internet censorship regulations and extensive firewall restrictions (Zou & Liu, 2024). All participants acknowledged that these access limitations created barriers to adoption, with many describing their engagement with global generative AIs as inconsistent or reliant on external resources.

For instance, two female participants majoring in philosophy and clinical medicine reported no prior use of ChatGPT due to their lack of access to VPNs or overseas accounts. In contrast, two male participants majoring in computer science stated that they had "legal access" to ChatGPT, facilitated by their internships in the AI industry and exchange studies in the United States. Participants who lacked formal access adopted various workaround strategies, including purchasing VPNs, borrowing accounts from friends or classmates and using intermediary services marketed through social media. However, these methods often proved unstable. For instance, Xiaoxue noted, "My ex-boyfriend was studying abroad, and we shared the same ChatGPT account. But recently, we broke up, so I can no longer use his account, meaning I no longer have access to ChatGPT." This account highlights how interpersonal factors can unexpectedly disrupt access, revealing a unique socio-technical dimension of *Facilitating Conditions* in censored environments. Moreover, participants expressed concerns about data privacy and service authenticity when relying on intermediaries. As Yuanfang questioned, "I know some of my classmates purchase intermediary services through social media platforms to access ChatGPT. However, I question whether they are truly connected to the real ChatGPT—nor do they know for certain." These cases underscore that *Facilitating Conditions* are not merely technical (e.g., internet access), but also social, economic, and relational.

In contrast to the constrained access to global AI, all participants reported prior experience utilizing domestic generative AI, such as including Kimi Chat (月之暗面), ChatGLM (智谱清言), Doubao (豆包), Wenxin Yiyan (文心一言), and Qwen (通义千问). These tools were readily accessible, often integrated with domestic platforms or apps (e.g., WeChat), thus lowering the barrier to entry. From the UTAUT2 perspective, this ease of access enhances both *Effort Expectancy* (minimal difficulty in learning or using the tool) and *Facilitating Conditions* (availability of technological infrastructure).

In addition, all participants reported that *Social Influence*—specifically, the generative AI choices made by their peers—positively affected their own decisions to use global or domestic generative models. Female students from non-STEM backgrounds were particularly likely to follow peers' recommendations. This aligns with UTAUT2's view that user behavior is influenced by perceived expectations from others.

### Paid Version and Free Version

Global generative AI platforms are generally priced around $20–$30 USD per month, offering access to advanced models like GPT-4, Claude 3, and Gemini 1.5 Pro. In contrast, domestic generative AI range from ¥49 to ¥199 CNY per month (approximately $6.74 to $27.38 USD), with similar tiered access.

The cost of technology can significantly influence users' adoption and utilization patterns. Venkatesh et al. (2012) highlight that "price value is positive when the benefits of using a technology are perceived to outweigh the monetary cost, and such price value has a positive impact on intention" (p.161). *Price Value* as a predictor of an individual's technology use behavior holds true in generative AI adoption in this study.

In total five participants subscribe to the paid version of global Generative AI, namely ChatGPT. They believe that the paid version (ChatGPT) provides significant benefits and the price is reasonable. As Maoxin explained, "Initially, I thought $20 [USD] was a bit expensive, but since I use it quite frequently and it helps improve my productivity, I find that spending $20 per month is actually acceptable." Similarly, Liqiu shared "The one I bought now costs ¥19.9 [Chinese currency, approximately $3 USD] a month to share an account with three other people. I think it is quite worth the money, because ChatGPT has really helped me a lot in my studies." Hedonic Motivation—the enjoyment or interest in using advanced AI—also influenced students' willingness to pay for generative AI, especially among STEM students. Several male students expressed enthusiasm for experimenting with powerful, cutting-edge models offered through global platforms, reinforcing the idea that interest and satisfaction contribute to technology adoption.

Interestingly, none of them pay for domestic Generative AI. Participants stated that the free version was already sufficient to meet their needs. As Yuheng explained, "I use the free version [domestic generative AI] because its existing functions already meet my basic needs." Similarly, Xunfei stated, "I believe it primarily depends on one's demand for the AI tool. In my case, my needs are relatively limited, so I do not place much importance on payment. I primarily use the free version."

*Price Value* is the primary factor in choosing between the paid and free versions of generative AI. However, *Effort Expectancy* also plays a role in the decision to pay for these services. Especially for the processing of the payment of global generative AI. As Qiji explained, "Making a payment is somewhat complicated. You need to use a foreign bank card, as domestic [Chinese] bank cards are not accepted. If you want to pay, the only option is to purchase it through Taobao, but I am not comfortable with this unofficial transaction method. So, even though I believe the paid version is better, I opted for the free one."

The additional effort required to navigate payment logistics deterred some students from subscribing, despite high interest or perceived value. This highlights how perceived complexity can reduce intention to adopt paid services, even when price and performance are favorable. Overall, students' decisions between paid and free versions of generative AI were shaped by a combination of Price Value, Effort Expectancy, and Hedonic Motivation, demonstrating the nuanced, multidimensional nature of technology adoption.

**Language: Chinese vs. English**

***Language Familiarity and Effort Expectancy***

Students demonstrate a notable level of strategic behavior when interacting with generative AI tools, particularly in terms of their choice between global and domestic platforms, and their use of Chinese versus English. These decisions are influenced by three main factors: language familiarity and habits, the nature of academic tasks, and perceived language capabilities of different AI models.

All 15 participants identified Chinese as their first and most familiar language, with English as a second language. Some also reported proficiency in additional languages such as Japanese, Spanish, or Latin. Language familiarity and usage habits influenced their engagement with generative AI. The majority (10 participants) used both Chinese and English, indicating a strong bilingual interaction pattern. Three participants primarily used Chinese, and only one relied exclusively on Chinese. One participant mainly used English, while none used English exclusively. This distribution is illustrated in Figure 2.

**Figure 2**

*Language Preferences in Generative AI Usage*

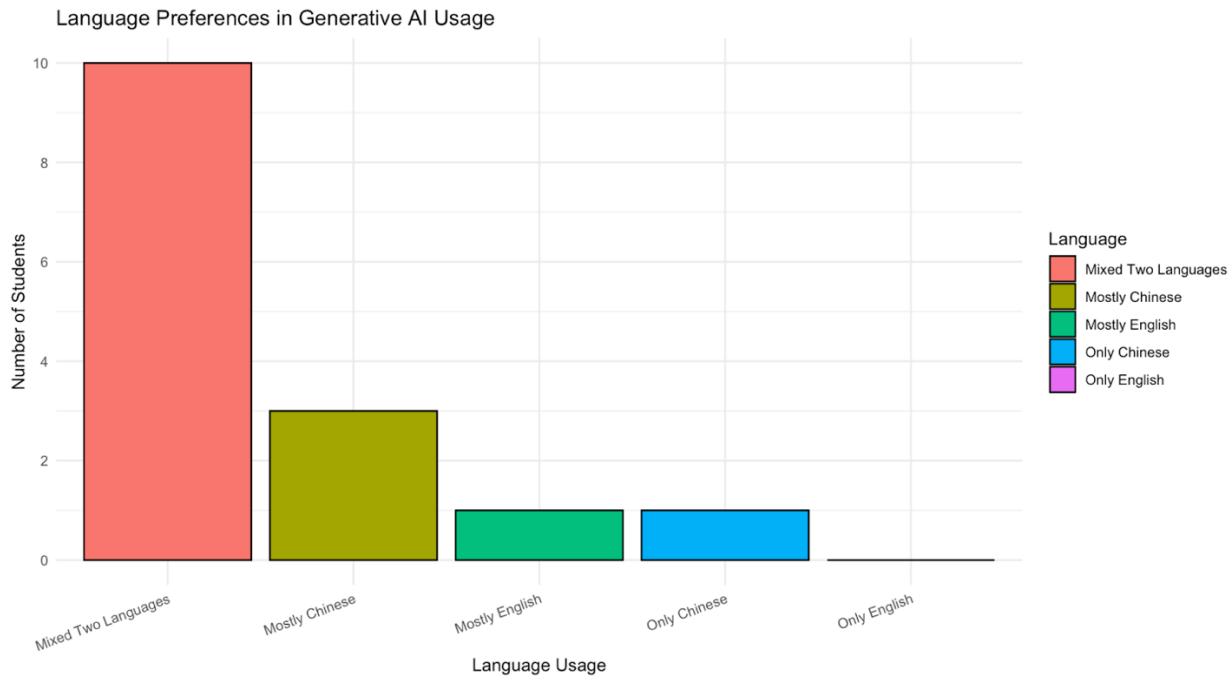

As Maoxin shared, "When brainstorming, I share my thoughts with generative AI, even with ChatGPT, in Chinese because it's more convenient. After all, it's my native language. When expressing abstract ideas, using Chinese often feels easier and more natural" Maoxin shared. As their mother tongue, Chinese offers students a sense of familiarity and comfort, making it the most convenient and accessible language for communication. This preference reflects students' comfort with Chinese, which provides lower *Effort Expectancy* and a stronger sense of fluency, especially in informal or creative interactions with generative AI.

### *Academic Tasks and Performance Expectancy*

Assignments content condition the preference of language and models. When the assignments are closely related to Western contexts and STEM disciplines, students are more likely to use English to interact with global generative AI. "It depends on the type of homework I'm doing. If it's English-related, such as programming or working with English documents, I naturally think in English and turn to ChatGPT first" Luyi shared. Mei also shared, "I use English with ChatGPT, since my core courses in Computer Science are primarily taught in English, and assignments are also completed in English. Additionally, coding is mostly done in English." Whereas, Yuheng expounds a different language preference, "Since I major in the humanities, my classes are conducted in Chinese, and most of the texts I study and use are Chinese literature. As a result, I primarily use Chinese and rely on domestic generative AI."

Students' perceptions of the language capabilities of different AI platforms also shape their usage behaviors. While existing evaluations show that the language gap between global and domestic models is narrowing, students consistently perceive global models such as ChatGPT as stronger in English, and domestic models as more capable in Chinese. As Xiyan describes, she primarily uses Chinese when interacting with domestic generative AI and switches to English when using ChatGPT. Similarly, Luyi explains:

> I don't ask questions in Chinese on ChatGPT. If I have a question in Chinese, I prefer to use domestic AI. In terms of English proficiency, domestic models are not as strong as ChatGPT. However, when it comes to Chinese, many domestic models, such as Kimi, outperform ChatGPT. In fact, we have conducted benchmark tests in both English and Chinese to confirm this.

Yuanfang agrees with Luyi's explanation, stating, "I use ChatGPT to translate English into Chinese, but I feel very disappointed. For example, it repeatedly uses the word 'express' (表达) throughout the entire article without variation. In contrast, domestic generative AI performs better in Chinese translation."

In summary, students adopt multilingual and cross-platform practices, adjusting their language and tool choices based on familiarity, task demands, and perceived model performance.

This strategic adaptability reflects their bilingual proficiency and the interplay between language habitus, academic background, and technical perception.

## Culture: Chinese Context vs. Western Context

### *Limitations of Global Generative AI in Traditional Chinese Contexts*

Higher education in China is deeply rooted in both the country's traditional cultural heritage and its contemporary political framework, specifically socialism with Chinese Characteristics (Zhu & Li, 2018). We first examine students' experiences using global and domestic generative AI when engaging with traditional Chinese culture in Chinese higher education. The following section will then explore the ideological context of socialism with Chinese characteristics.

Traditional Chinese culture, rooted in philosophies like Confucianism and Taoism, encompasses art, festivals, medicine, and values such as harmony and family, forming a rich heritage that continues to influence and inspire Chinese higher education (Gu, 2006). This cultural foundation is deeply embedded in courses such as traditional Chinese literature, Chinese philosophy, and Chinese painting (Zhang, 2024). It is therefore unsurprising that students increasingly rely on generative AI for support in completing assignments related to traditional Chinese culture.

Based on participants' responses, global generative AI seems to struggle with accurately interpreting traditional Chinese culture. As Qiji shared, "I asked ChatGPT to help with my homework on Chinese philosophy, but it had no real understanding of key concepts such as 'Tao' (道), 'Benevolence' (仁), and 'Filial Piety' (孝)." Similarly, Zhaolan illustrates:

Blank space is a distinctive feature of Chinese painting, reflecting Taoist philosophy, traditional aesthetics, and national identity. Unlike the subdued elegance of Chinese art, I found AI-generated works from ChatGPT and Midjourney disappointing—they resembled Western oil paintings, lacked blank space, and used overly bright colors, despite my detailed prompts.

Conversely, since domestic generative AIs are trained on a larger dataset with Chinese contextual information, they have a better understanding of content related to traditional Chinese culture (Zhang et al., 2021). "Domestic generative AI often creates overly scholarly content for traditional Chinese culture assignments. For example, when describing a Fujian ancient building, the output felt excessively refined and antique, as if written by someone deeply versed in classical literature" Xunfei shared. Students tend to rely more on domestic generative AI for assignments or learning related to traditional Chinese culture.

### Strengths of Domestic Generative AI in Socialism with Chinese Characteristics

The ideology on Socialism with Chinese Characteristics embedded in higher education curriculum. Where all Chinese college students are required to take ideological courses, such as Maoism and Marxism (Wang, 2024). These courses typically involve assignments that require students to submit reflection reports on ideological topics. Within this academic context, global generative AI often struggles to meet students' *performance expectancy*. As Xiaoxue shared, "When I used GPT for areas that strongly reflect socialism with Chinese characteristics, the output often felt somewhat off. It carried a distinctly Western tone that did not align with the intended context."

In contrast, domestic generative AI demonstrates superior performance compared to global generative AI when handling tasks related to Socialism with Chinese Characteristics. As Yuyan highlights, "When working on assignments related to Socialism with Chinese Characteristics, I find that using domestic generative AI is much more convenient."

Although domestic generative AI performs better in handling tasks related to Socialism with Chinese Characteristics, they come with a notable limitation. These models are programmed with predefined restrictions on sensitive and censored terms, making them incapable of processing tasks that involve such content. As Yalin shared, "While working on my homework, I encountered certain politically sensitive words, and domestic generative AI would either indicate that they were unable to generate content containing those terms or state that they could not address the topic altogether."

Accordingly, students are more likely to use domestic generative AI for assignments or learning related to traditional Chinese culture and socialism with Chinese characteristics due to the varying performance expectancy of different models.

## Attitude:  Human vs. Machine

### *"Please / Thank you" or "Do it"*

Generative AI is reshaping behavioral modeling and data analysis by producing realistic data, text, and images (Chen et al., 2024). But how do students perceive this transformative technology? Is it simply an advanced machine, or something more (Xie, 2024)? In the semi-structured interviews, one key question probed students' perceptions of generative AI. Interestingly, participants did not distinguish between global and domestic, instead grouping them under the broad label of "Generative AI." Six students described it purely as a machine, four viewed it as both a machine and an assistant, two saw it solely as an assistant, another two considered it a friend, and one participant described it as a nurturing co-learner.

Due to the current limitations of generative AI (e.g., voice chat lag), some participants perceive it as merely a machine. However, they acknowledge that this is their present view and that their perspective may evolve in the future, potentially seeing generative AI as more than just a machine. The role of generative AI is fluid and continuously shifting.

Some participants have a more nuanced perspective. For example, Liqiu reflects, "I see it as a tool, but sometimes I feel a bit guilty, especially when generative AI does a lot of work for me. However, at the same time, I also feel there's no reason to feel guilty about using a machine." This sentiment is not uncommon. Mei shares a similar experience, stating, "I tend to view generative AI purely as a tool. However, I sometimes chat with ChatGPT, and occasionally, its responses feel very human, making me unconsciously perceive it as kind and pleasant."

Unconscious thought patterns are reflected in unconscious (and sometimes conscious) behaviors, such as using polite expressions when interacting with generative AI. Two participants explicitly stated that they never use honorific expressions and instead give direct commands, such as "do this… do that…" However, the remaining participants acknowledged that they either consciously or unconsciously use polite expressions when engaging with generative AI. Xiyan notes:

> "One thing I find really funny is that, I don't know why every time I ask [generative AI] a question, I always say 'please' and 'thank you'…One day when I was writing, I discovered this phenomenon. I felt very strange and I inquired, "why should I say 'please' to you [generative AI]?"

Xiyan's question received various answers, including improved AI performance, a sense of ritual, polite habits, and natural response. Maoxin shared: "Showing respect might make it perform better. When I need its help, I sometimes even beg. It's a psychological action and a fun aspect of research." Sijia observed: "Unlike using Baidu in the past, simply opening the conversation page [of generative AI] makes me feel the need for a more formal and ritualistic interaction." Yuyan's response had a science-fiction twist: "Just like in The Machine, when the robots rebelled, I survived because I always said 'please' to them. They spared me because I was kind to them."

Whether generative AI, both global and domestic, is perceived as anthropomorphic intelligence or merely a cold machine is a subjective question. Students have varying interpretations of generative AI and different reasons for saying "please" during their interactions.

### *Confidence or Anxiety*

The terms confidence and anxiety refer to students' confidence levels regarding the impact of both global and domestic generative AI on their studies and personal development. Specifically, studies refer to students' academic performance, personal development related to job opportunities. In other words, here we explore how the students view the relationship between themselves and generative AI in higher education contexts.

The confidence level in academic performance through the integration of generative AI is examined by dividing academic performance into two aspects: score improvement and learning improvement. The confidence levels in these two areas vary. The majority of participants, approximately 11 individuals, lack confidence in score improvement through the use of both global and domestic generative AI. However, participants are generally more confident that generative AI enhances learning efficiency, particularly in terms of time-saving benefits.

The confidence level of job opportunities under global and domestic generative AI emerging context. Participants expressed more anxiety than confidence regarding job opportunities in the era of generative AI, though the level of anxiety varied among individuals.

Sijia shared, "It does influence my career choice to some extent. Now, I have to prepare for a job that is less likely to be replaced." Similarly, Yalin reflected, "If I become a doctor in the future and AI becomes advanced enough, it could handle tasks like analyzing medical images or diagnosing conditions. I believe this would impact my employment." The anxiety does not stem solely from the advancement of technology but also from the increasing competition among technologists. Luyi remarked, "There is definitely some pressure [on job opportunities]. It's not that AI itself poses a threat, but rather the intense competition among the many developers in the field."

One intriguing phenomenon is that STEM students tend to worry more about job opportunities for Social Science and Humanities students, while Social Science and Humanities students are more concerned about job opportunities for STEM students. "It appears that programmers may indeed face unemployment. While AI may become highly intelligent, it cannot replace humanistic care. Therefore, society will continue to need students in the humanities and social sciences" Yuheng states. Conversely, Luyi argues, "generative AI is likely to replace many liberal arts students. Meanwhile, STEM majors have little to worry about for now. After all, ChatGPT is not yet advanced enough to threaten their jobs."

While individual students express concern about their career prospects amid rapid global and domestic expansion of generative AI, collective perspectives reveal notable optimism. Both STEM and Social Science/Humanities students demonstrate greater confidence in their respective job markets when considering field-specific opportunities. This dichotomy suggests that students perceive stronger sectoral buffers against AI-driven labor market disruptions when evaluating their disciplines collectively rather than personally. This finding is surprising, as it challenges the common research assumption that STEM students benefit more in the AI era (Corrales-Herrero, & Rodríguez-Prado, 2024; Okrent & Burke, 2021).

**Discussion**

**The Assurance of Equitable Access to Free and Open-source Generative AI**

In today's rapidly evolving digital landscape, the rise of generative AI is transforming how we communicate, work, and create. However, this technological revolution also underscores a critical yet often overlooked issue: the widening digital divide (Hendawy, 2024). The digital divide stems from multiple factors. Political constraints can hinder students from countries such as China, Russia, and Iran from accessing global generative AI. Economic barriers, including subscription fees and high monthly costs, further restrict students' access to advanced versions of generative AI, particularly on a global scale. In response, some students resort to sharing accounts or navigating legal grey areas to gain access, practices that not only raise privacy concerns but also exacerbate the digital gap. Additionally, infrastructure limitations play a role, as certain communities face restricted access due to domain limitations and internet censorship (Ragnedda & Ragnedda, 2020). The challenges Chinese students face in accessing global generative AI are not isolated but may extend to students in similar contexts worldwide. To foster equitable and responsible learning environments, ensuring open and unrestricted access to generative AI for all students is essential.

**The Imperative for Non-English Generative AI**

Generative AI faces a significant language challenge, as English remains the dominant language in its development and application (Choudhury, 2023). The widespread use of English in generative AI can marginalize non-English languages, limiting access to educational content for indigenous language speakers and potentially hindering their ability to learn in their native tongue (Nyaaba et al., 2024). This study finds that domestic generative AI outperforms global generative

AI in the Chinese language, providing students with viable alternatives to integrate domestic AI tools into their learning processes. However, China and Silicon Valley are two leading AI superpowers (Lee, 2018). In contrast, many non-English-speaking nations lack the technological and economic resources to develop generative AI tailored to their linguistic needs. As Nyaaba et al. (2024) warn, "this may inadvertently marginalize indigenous and minority languages, potentially accelerating language erosion and the dilution of cultural heritage among certain minoritized groups" (p. 9). To harness generative AI for inclusive societal transformation in higher education, it is essential to invest in and enhance non-English generative AI at both global and domestic levels.

**The Necessity of Generative AI with Local Cultural Awareness**

Global generative AI encounters difficulty in understanding traditional Chinese culture and Socialism with Chinese Characteristics in this study, similarly in West Africa, Central America, Southeast Asia, and other contexts (Nyaaba et al., 2024). Generative AI primarily trained on data from Western sources may unintentionally emphasize Western perspectives, values, and ideologies (Brand, 2023). Scholars have critiqued generative AI as a tool of "AI Empire" (Tacheva & Ramasubramanian, 2023) and "Digital Neocolonialism" (Arora et al., 2023), highlighting concerns over its cultural biases. These biases can exclude students from diverse cultural backgrounds by providing examples that fail to align with their lived experiences and environments (Mollema, 2024). This form of cultural dominance risks marginalizing non-Western perspectives in educational content, ultimately diminishing its relevance and effectiveness for students from underrepresented backgrounds. To address this issue and decolonize the cultural imperialism embedded in generative AI within educational contexts, it is essential to enhance Generative AI's awareness and integration of local cultural perspectives.

**The Need for a Future-oriented Understanding of Generative AI**

Is generative AI merely a tool, or does it transcend that role? Why do students, both consciously and unconsciously, say "thank you" and "please" when interacting with generative AI? This is not a simple question, and it warrants deeper exploration. The interaction between students and generative AI, as well as the broader human-machine relationship, is fluid, evolving, and constantly shifting. To fully grasp the role of generative AI in higher education, scholars, educators, and policymakers must adopt a human-centric approach (Holmes & Miao, 2023), as emphasized by UNESCO and other educational institutions (U.S. Department of Education, 2023). At the same time, it is essential to remain attentive to the complex and evolving nature of generative AI's integration into students' learning experiences, particularly across different disciplines and gender identities.

**Conclusion**

This study illuminates how Chinese undergraduates navigate both global and domestic generative AI in their learning, revealing four key shaping factors: access limitations, linguistic capabilities, and cultural familiarity. Students strategically select generative AI based on perceived usefulness, favoring global AI for advanced functionality yet facing access barriers that spur ethically questionable workarounds (e.g., VPNs, account sharing). A clear dichotomy emerges: students proficient in English and Western contexts engage more with global AI, while others prefer domestic tools for superior Chinese-language processing and cultural alignment. However, domestic AI's content filtering and limited outputs constrain academic exploration.

These findings advance three critical discussions: (1) the political-linguistic dimensions of AI adoption in education, (2) digital equity in non-Western contexts, and (3) cultural diversity in technology design. They underscore the urgency of developing human-centered, context-sensitive AI ecosystems that balance functionality with inclusivity.

While this single-site study offers rich qualitative insights, future research should employ comparative mixed methods across institutions to strengthen generalizability. As generative AI reshapes global education, such work will be vital to ensure equitable learning environments that respect linguistic, cultural, and epistemological diversity.